\documentclass[a4paper,11pt]{article}
\usepackage{pos}
\usepackage{setspace}
\usepackage{wrapfig}
\usepackage{lipsum}

\usepackage{cleveref}
\Crefformat{equation}{Eq.~(#2#1#3)}
\Crefmultiformat{equation}{Eqs.~(#2#1#3)}{ and~(#2#1#3)}{, (#2#1#3)}{ and~(#2#1#3)}
\Crefformat{figure}{Fig.~#2#1#3}
\Crefformat{section}{Sec.~#2#1#3}

\allowdisplaybreaks

\def\latmom{\left( \frac{2\pi}{L} \right)}

\newcommand{\bv}[1]{{\bf{#1}}}

\newcommand{\qb}{\bv{q}}
\newcommand{\pb}{\bv{p}}

\def\llangle{\left\langle}
\def\rrangle{\right\rangle}
\newcommand{\ea}[1]{{\llangle #1 \rrangle}}
\newcommand{\Dfb}{\overleftrightarrow{D}}

\title{Towards nucleon structure function moments and parton momentum fractions from lattice QCD}
\ShortTitle{Towards nucleon structure function moments and \dots}

\author*[a]{K.~U.~Can}
\author[b]{R.~Horsley}
\author[c]{P.~E.~L.~Rakow}
\author[d]{G.~Schierholz}
\author[e]{H.~St\"{u}ben}
\author[a]{R.~D.~Young}
\author{J.~M.~Zanotti${^a}$ (QCDSF Collaboration)}

\affiliation[a]{CSSM, Department of Physics, The University of Adelaide,
Adelaide SA 5005, Australia.}
\affiliation[b]{School of Physics and Astronomy, University of Edinburgh, Edinburgh EH9 3JZ, UK.}
\affiliation[c]{Theoretical Physics Division, Department of Mathematical Sciences, University of Liverpool, Liverpool L69 3BX, UK.}
\affiliation[d]{Deutsches Elektronen-Synchrotron DESY, Notkestr. 85, 22607 Hamburg, Germany.}
\affiliation[e]{Regionales Rechenzentrum, Universit\"{a}t Hamburg, 20146 Hamburg, Germany.}

\emailAdd{kadirutku.can@adelaide.edu.au}

\abstract{
We calculate the lowest even isovector moment of the $F_2$ structure function in $2+1$-flavour lattice QCD with varying quark masses corresponding to $m_\pi \approx [410, 360, 300] \; {\rm MeV}$, at a fixed volume of $V = 48^3 \times 96$ and coupling $\beta = 5.65$ ($a = 0.068(3) \, {\rm fm}$). We directly compute the physical Compton amplitude using the Feynman-Hellmann approach and extract moments of the physical structure function. We report on the quark-mass dependence of the lowest isovector moment and estimate its value at the physical quark-mass point with $\sim 10\%$ uncertainty at fixed $Q^2$. By analysing the $Q^2$ dependence of the moments at the $SU(3)$ symmetric point ($m_\pi \approx 410 \; {\rm MeV}$), we separate the leading- and higher-twist contributions and estimate the parton momentum fraction, $\langle x \rangle_{u-d}$, which agrees with existing results.
}

\FullConference{The XVIth Quark Confinement and the Hadron Spectrum Conference (QCHSC24)\\
 19-24 August, 2024\\
 Cairns Convention Centre, Cairns, Queensland, Australia\\}

\begin{document}
\maketitle

\section{Introduction}
Decoding the internal structure of hadrons from first principles is a fundamental challenge in particle and nuclear physics, with implications for high-energy, nuclear, and astroparticle physics. The properties of hadrons, from the hybrid structure of quark and meson degrees of freedom at low energies to their partonic structure at short distances, are encapsulated in structure functions. Lattice QCD provides a first-principles framework for computing these functions, offering a non-perturbative approach to hadron structure.

The connection between nucleon structure functions and quark dynamics is described using the parton model. While this model provides an intuitive interpretation of deep inelastic scattering (DIS) data, it assumes that partons interact elastically and incoherently with the incoming lepton. The operator product expansion (OPE) for the Compton amplitude systematically classifies contributions by twist, with the parton model capturing only twist-two terms, neglecting power corrections from higher-twist operators. However, renormalization and operator mixing inherently link higher-twist contributions with leading-twist terms \cite{Martinelli:1996pk,Beneke:1998ui}, making their separation non-trivial.

Lattice QCD calculations of nucleon structure functions have traditionally focused on leading-twist matrix elements, particularly the lowest moments of parton distribution functions (PDFs), building on work of~\cite{Martinelli:1988rr,Gockeler:1995wg,Gockeler:1996mu,Gockeler:1998ye}. Most notably, the quasi-~\cite{Ji:2013dva} and pseudo-PDF~\cite{Radyushkin:2017cyf} approaches have made significant contributions to advancing the calculations. We refer the reader to the reviews~\cite{LIN2018107,FLAG:2024oxs} that collect and summarise the lattice QCD community's efforts in calculating the $x$-dependent PDFs and PDF moments. These approaches, however, suffer from operator mixing with higher-twist contributions, leading to power-divergent renormalization effects that require nonperturbative subtraction \cite{Martinelli:1996pk,Rossi:2017muf,Braun:2018brg}. 

An alternative is the direct lattice calculation of the forward Compton amplitude using the Feynman-Hellmann approach~\cite{Chambers:2017dov}, which circumvents issues of operator mixing and renormalization inherent in the OPE. Its application to the calculations of the moments of unpolarised~\cite{PhysRevD.102.114505,QCDSFUKQCDCSSM:2022ncb, Can:2025zsr} and polarised~\cite{Can:2022chd} nucleon structure functions, and generalised parton distributions~\cite{Alec:2021lkf,Hannaford-Gunn:2024aix}, have proven the effectiveness of this method. This strategy does not require an explicit separation of twist contributions, allowing for a more complete determination of nucleon structure functions, including power corrections. Furthermore, by mapping the OPE onto the Compton amplitude, it is possible to obtain renormalized Wilson coefficients and operator matrix elements in a self-consistent manner.

In this contribution, extending upon Ref.~\cite{QCDSFUKQCDCSSM:2022ncb}, we report on our calculation of the forward Compton amplitude of a nucleon away from the $SU(3)$ symmetric point at a fixed $Q^2$ to quantify its quark-mass dependence, and extract the lowest isovector moment of the $F_2$ structure function, i.e. the isovector momentum fraction. Additionally, we reanalyse the physical structure function moments that have been calculated in Ref.~\cite{QCDSFUKQCDCSSM:2022ncb} and by studying their $Q^2$ dependence we extract the lowest even isovector PDF moment, $\langle x \rangle_{u-d}$.  

The rest of this contribution is organized as follows: formal definitions of the Compton amplitude and the structure functions, along with the connection between the OPE and the dispersion relation are given in \Cref{sec:ca}. We summarise the key points of the second order Feynman-Hellmann theorem in \Cref{sec:fh}. Our lattice setup, along with the details of our calculations and analysis are given in \Cref{sec:simu}. Results for the lowest isovector moment of the $F_2$ structure function are presented in \Cref{sec:res}. We summarize our findings and provide an outlook in \Cref{sec:sum}.

\section{Compton amplitude and moments of structure functions} \label{sec:ca}
In order to access the structure functions, we consider the unpolarised forward Compton tensor, defined as the time-ordered product of two electromagnetic currents,
\begin{equation} \label{eq:compamp}
  T_{\mu\nu}(p,q) = i \int d^4z\, e^{i q \cdot z} \rho_{s s^\prime} \ea{p,s^\prime \left| \mathcal{T}\left\{ \mathcal{J}_\mu(z) \mathcal{J}_\nu(0) \right\} \right|p,s},
\end{equation}
which is decomposed in terms of Lorentz-invariant functions,
\begin{align}\label{eq:comptensor}
  \begin{split}
    T_{\mu\nu}(p,q) &= \left( -g_{\mu\nu} + \frac{q_\mu q_\nu}{q^2} \right) \mathcal{F}_1(\omega,Q^2)
    + \frac{\hat{P}_\mu \hat{P}_\nu}{p \cdot q} \mathcal{F}_2(\omega,Q^2),
  \end{split}
\end{align}
where $q$ ($p$) is the momentum of the virtual photon (nucleon), $\hat{P}_\mu \equiv p_\mu - (p \cdot q) q_\mu/q^2$, $\omega = (2p \cdot q)/Q^2$ and $Q^2 = -q^2$. The Lorentz invariant Compton structure functions $\mathcal{F}_{1,2}$ are related to the physical, $x$-dependent structure functions $F_{1,2}$ via the optical theorem, $\operatorname{Im}\mathcal{F}_{1,2}(\omega,Q^2) = 2\pi F_{1,2}(x,Q^2)$. Utilizing analyticity, crossing symmetry, and the optical theorem, the Compton structure functions satisfy the dispersion relations~\cite{Drechsel:2002ar},
\begin{align} 
  \label{eq:compdisp1}
  \overline{\mathcal{F}}_1(\omega,Q^2) &= 2\omega^2 \int_0^1 dx \frac{2x \, F_1(x,Q^2)}{1-x^2\omega^2-i\epsilon}, \\
  \label{eq:compdisp2}
  \mathcal{F}_2(\omega,Q^2) &= 4\omega \int_{0}^1 dx\, \frac{F_2(x,Q^2)}{1-x^2\omega^2-i\epsilon},
\end{align}
where $\overline{\mathcal{F}}_1(\omega,Q^2) = \mathcal{F}_1(\omega,Q^2)-\mathcal{F}_1(0,Q^2)$ is the once-subtracted structure function. Expanding the integrand in \Cref{eq:compdisp2} around $\omega=0$, we express the Compton structure functions as infinite sums over the Mellin moments of the inelastic structure functions,
\begin{align} 
  \label{eq:ope_moments1}
  \overline{\mathcal{F}}_{1}(\omega,Q^2) &=\sum_{n=0}^\infty 2\omega^{2n} M^{(1)}_{2n}(Q^2), \\
  \label{eq:ope_moments2}
  \frac{\mathcal{F}_2(\omega,Q^2)}{\omega} &= \sum_{n=1}^\infty 4\omega^{2n-2} M^{(2)}_{2n}(Q^2),
\end{align}
where,
\begin{align} 
  \label{eq:moments1}
  M^{(1)}_{2n}(Q^2) &= 2\int_0^1 dx\, x^{2n-1} F_1(x,Q^2), \\
  \label{eq:moments2}
  M^{(2)}_{2n}(Q^2) &= \int_{0}^1 dx\,x^{2n-2} F_{2}(x,Q^2),
\end{align}
for $n = 1, 2, \dots$. Previous studies extracting the lowest even Mellin moment of $F_1$ required a polynomial fit in $\omega$~\cite{PhysRevD.102.114505}. However, the lowest even Mellin moment of $F_2$, $M_2^{(2)}(Q^2)$, is directly accessible at $\omega = 0$, corresponding to a nucleon at rest. As a result, it can be determined directly from the Compton amplitude without the need for a polynomial fit in $\omega$, since the right-hand side of \Cref{eq:ope_moments2} remains well-defined at $\omega = 0$.

Within the framework of the current lattice calculation, the Compton tensor can be evaluated for multiple values of $\omega$, allowing for the extraction of structure function moments. Treating $Q^2$ as an external scale ensures a direct connection to the relevant physical amplitudes. Although these lattice calculations yield the physical moments $M_{2n}^{(2)}$, we emphasise that in the limit of asymptotically large $Q^2$, the leading-twist contributions become dominant. In this regime, the moments are primarily governed by the well-established parton distribution functions (PDFs), $v_{2n}$,
\begin{equation} \label{eq:leading-twist}
  M^{(j)}_{2n,f}(Q^2) = C_{2n,f}^{(j)}\left(\frac{Q^2}{\mu^2},g(\mu)\right)v_{2n}^f(\mu)+\mathcal{O}\left(\frac{1}{Q^2}\right), \quad j=1,2
\end{equation}
where $f$ denotes the partonic flavour. The short distance structure of the operator product in \Cref{eq:compamp} is encoded in the Wilson coefficients, $C$, given by (to leading order):
\begin{equation}
  C_{2n,f}^{(j)}=\mathcal{Q}_f^2+\mathcal{O}(g^2), \quad j=1,2
\end{equation}
where $\mathcal{Q}_f$ is the electric charge of the parton, and the long-distance hadronic features are encoded in the matrix elements of local operators, $v$, renormalized at some scale $\mu$, defined by
\begin{equation}
  \langle p, s|\left[\mathcal{O}_f^{\{\mu_1\ldots \mu_n\}}-\operatorname{Tr}\right] |p,s\rangle =
  2v_n^f\left[p^{\mu_1}\ldots p^{\mu_n}-\operatorname{Tr} \right],
\end{equation}
in terms of the traceless and symmetric parts of the local quark bilinears:
\begin{equation}
  \mathcal{O}^{\{\mu_1\ldots\mu_n\}}_q=i^{n-1}\overline{\psi}_q\gamma^{\mu_1}\Dfb^{\mu_2}\cdots \Dfb^{\mu_n}\psi_q,
\end{equation}
and similarly for the gluons
\begin{equation}
  \mathcal{O}^{\{\mu_1\ldots\mu_n\}}_g=i^{n-2}\operatorname{Tr} F^{\mu_1\nu}\Dfb^{\mu_2}\cdots \Dfb^{\mu_n} F^{\mu_n}_{\phantom{\mu_n}\nu},
\end{equation}
where $\Dfb=\tfrac12(\overrightarrow{D}-\overleftarrow{D})$.

In the following discussion, we provide the details of our procedure for extracting the physical structure function moments directly from the Compton amplitude obtained via a lattice QCD calculation. Furthermore, we show how the PDF moments (the leading-twist matrix elements) can be estimated via studying the $Q^2$ dependence of the moments of the physical structure functions.

\section{The Feynman-Hellmann approach} \label{sec:fh}
We compute the Compton amplitude by means of the second-order Feynman-Hellmann theorem as derived and described in detail in~\cite{PhysRevD.102.114505}. Here we summarise the procedure relevant to this work. We perturb the fermion action by the vector current,
\begin{equation}\label{eq:fh_perturb}
    S(\lambda) = S_0 + \lambda \int d^4z (e^{i \bv{q} \cdot \bv{z}} + e^{-i \bv{q} \cdot \bv{z}}) \mathcal{J}_{\mu}(z),
\end{equation}
where $S_0$ is the unperturbed action, $\lambda$ is the strength of the coupling between the quarks and the external field, $\mathcal{J}_{\mu}(z) = Z_V \bar{q}(z) \gamma_\mu q(z)$ is the electromagnetic current coupling to the quarks, $q=(0,\qb)$ is the external momentum inserted by the currents and $q_0=0$ by construction in the Feynman-Hellmann approach~\cite{PhysRevD.102.114505}. $Z_V$ is the renormalisation constant for the local electromagnetic current, which has been determined in Ref~\cite{Constantinou:2014fka}. The perturbation is introduced on the valence quarks only, hence only quark-line connected contributions are taken into account in this work. For the perturbation of valence and sea quarks see~\cite{Chambers2015}.

The Feynman-Hellmann relation between the second-order energy shift and the Compton amplitude is given by~\cite{PhysRevD.102.114505},
\begin{equation} \label{eq:secondorder_fh}
    \left. \frac{\partial^2 E_{N_\lambda}(\pb, \qb)}{\partial \lambda^2} \right|_{\lambda=0} 
    = -\frac{T_{\mu\mu}(p,q) + T_{\mu\mu}(p,-q)}{2 E_{N}(\pb)},
\end{equation}
where $T_{\mu\nu}$ is the Compton tensor defined in \Cref{eq:comptensor}, and $E_{N_\lambda}(\bv{p})$ is the nucleon energy at momentum $\bv{p}$ in the presence of a background field of strength $\lambda$. This expression is the principal relation that we use to access the Compton amplitude and hence the Compton structure functions.

\section{Calculation and analysis details} \label{sec:simu}
Our calculations are performed on QCDSF's 2$+$1-flavour gauge configurations generated with a stout-smeared non-perturbatively $\mathcal{O}(a)$-improved Wilson action for the dynamical up/down and strange quarks and a tree-level Symanzik improved gauge action~\cite{Cundy:2009yy}. We utilise three ensembles with varying quark masses corresponding to $m_\pi \approx [410, 360, 300] \; {\rm MeV}$, at fixed volume $V = 48^3 \times 96$ and coupling $\beta = 5.65$ corresponding to a lattice spacing $a = 0.068(3) \, {\rm fm}$. The quark-mass trajectory starts from the $SU(3)$ symmetric point $(m_\pi \approx 410 \; {\rm MeV})$, where the masses of all three quark flavours are set to approximately the physical flavour-singlet mass $\overline{m} = (2 m_s + m_l)/3$~\cite{Bietenholz:2010jr,Bietenholz:2011qq}, and approaches the physical point along the $\overline{m} = {\rm constant}$ line. 

In this work, we perform two analyses focusing on the lowest isovector moment of the $F_2$ structure function. The first analysis quantifies the quark-mass dependence of the moment at fixed $Q^2 \sim 5 \; {\rm GeV}^2$. For this analysis, we perform $\mathcal{O}(10^3)$ measurements by employing two sources with randomly chosen locations on each of the three ensembles of size $\mathcal{O}(500)$ gauge configurations. Quark fields are smeared in a gauge-invariant manner by Jacobi smearing~\cite{Allton:1993wc}, where the smearing parameters are tuned to produce a rms radius of $\simeq 0.5$ fm for the nucleon. 

In the second analysis, we study the $Q^2$ dependence of the lowest isovector moment at the $SU(3)$ symmetric point to extract the leading-twist part and determine the PDF moment, and additionally, estimate the leading higher-twist contribution. We work in the range $1 \lesssim Q^2 \lesssim 7.5 \; {\rm GeV}^2$ and utilise the correlators calculated in a previous work~\cite{QCDSFUKQCDCSSM:2022ncb}.

As discussed in \Cref{sec:fh}, we access the Compton structure functions through the energy shifts. Our procedure to calculate the energy shifts and extract the Compton amplitude has been laid out in Refs.~\cite{PhysRevD.102.114505,QCDSFUKQCDCSSM:2022ncb,Can:2025zsr}. Here we summarise the key points which underlie the both analyses mentioned above. In order to isolate the $\mathcal{F}_2$ Compton structure function from the Compton tensor in \Cref{eq:comptensor}, we set $q_3=p_3=0$ and consider the current components $\mathcal{J}_{0}$ and $\mathcal{J}_{3}$. Working in Minkowski space and noting $q_0=0$ in the Feynman-Hellmann approach, we have,
\begin{equation} \label{eq:compF2}
    \frac{\mathcal{F}_2(\omega,Q^2)}{\omega} = \frac{Q^2}{2 E_N^2} \left[ T_{00}(p,q) + T_{33}(p,q) \right],
\end{equation}
where the RHS is well-defined at $\omega=0$. 

Our calculations are done for several values of $\bv{q}$. Multiple values of $\omega$ are accessed by varying the nucleon momentum $\bv{p}$ for a fixed $\bv{q}$. A list of $\omega$ values used in the analysis is provided in Ref.~\cite{QCDSFUKQCDCSSM:2022ncb}. By attaching the current selectively to the $u$ and $d$ quarks, respectively, we obtain the flavour diagonal $uu$ and $dd$ contributions. The asymptotic behaviour of the perturbed two-point correlator at large Euclidean time takes the familiar form,
\begin{equation} \label{eq:G2spec}
      G^{(2)}_{\lambda}(\pb,\qb,t) \simeq A_\lambda(\pb,\qb) e^{ - E_{N_{\lambda}}(\pb,\qb) \, t },
\end{equation}
where $E_{N_{\lambda}}$ is the perturbed energy of the ground state nucleon in the presence of an external field and $A_\lambda$ the corresponding overlap factor. In order to extract the second-order energy shift, we calculate the following combination of perturbed energies,
\begin{align}
    \Delta E_{N_{\lambda}}(\pb,\qb) &= \frac{1}{2} \left[
    E_{N_{+\lambda}}(\pb,\qb) +
    E_{N_{-\lambda}}(\pb,\qb) - 2E_N(\pb) \right] \\
    \label{eq:enshift_oo}
    &= \frac{\lambda^2}{2!} \left. \frac{\partial^2 E_{N_{\lambda}}(\pb,\qb)}{\partial \lambda^2} \right|_{\lambda=0} + \mathcal{O}(\lambda^4),
\end{align}
which isolates the desired term (\Cref{eq:enshift_oo}) appearing on the LHS of \Cref{eq:secondorder_fh}. By forming the following ratio of perturbed to unperturbed correlators,
\begin{equation} \label{eq:ratio_fd}
    \mathcal{R}^{qq}_{\lambda}(\pb,\qb, t) \equiv \frac{G^{(2)}_{+\lambda}(\pb,\qb, t) G^{(2)}_{-\lambda}(\pb,\qb, t)}{\left( G^{(2)}(\pb, t) \right)^2} \xrightarrow{t \gg 0} \tilde{A}_{\lambda}^{qq} e^{-2\Delta E^{qq}_{N_\lambda}(\pb,\qb) \, t},
\end{equation} 
the second-order energy shift is extracted, where $qq = uu, \, dd$ and this ratio isolates the energy shift ($\Delta E_{N_\lambda}^{qq}(\pb,\qb)$) only at even orders of $\lambda$. Here, $G^{(2)}(\pb, t)$ is the unperturbed correlator $G^{(2)}(\pb,t) \simeq A(\pb) e^{ - E_N(\pb) \, t }$, and we have collected all overlap factors into a single constant $\tilde{A}_\lambda(\pb,\qb)$ which is irrelevant for our discussion of energy shifts.

We proceed with a weighted averaging method outlined in~\cite{NPLQCD:2020ozd} and applied to Compton amplitude in~\cite{Can:2025zsr} to extract the energy shifts from the ratio defined in \Cref{eq:ratio_fd}. Our analysis here closely follows that of Ref.~\cite{Can:2025zsr}, however, we sketch the procedure below for completeness. To determine an energy shift, we perform several single exponential, \Cref{eq:G2spec}, fits to the ratio, \Cref{eq:ratio_fd}, with fixed $t_{\rm max}$ and varying $t_{\rm min}$, where we demand $t_{\rm max}-t_{\rm min} \ge 3a$. The weight on each fit is calculated by~\cite{NPLQCD:2020ozd},
\begin{equation} \label{eq:weight}
  w^f = \frac{p_f \, (\delta \mathcal{O}^f)^{-2}}{\sum_{f^\prime} p_{f^\prime} (\delta \mathcal{O}^{f^\prime})^{-2}},
\end{equation} 
where $p_f$ is the two-sided p-value of a $\chi^2$ distribution. Here $f$ denotes the choice of fit window, and $\mathcal{O}$ is an unbiased estimator of a generic quantity, e.g. the energy shift extracted from the ratio. $\delta \mathcal{O}^f$ is the statistical uncertainty on $\mathcal{O}^f$ estimated by a bootstrap analysis. Lastly, the final estimate of the quantity of interest, $\bar{\mathcal{O}}$, and its uncertainty, $\delta \bar{\mathcal{O}}$ are calculated via,
\begin{subequations}
  \noindent\centering
  \begin{minipage}{0.48\textwidth}
    \begin{align}
      \label{eq:wavg}
      \bar{\mathcal{O}} &= \sum_f w^f \mathcal{O}^f, \\ 
      \label{eq:wstat}
      \delta_{\rm stat} \bar{\mathcal{O}}^2 &= \sum_f w^f (\delta \bar{\mathcal{O}}^f)^2,
    \end{align}
  \end{minipage}
  \hfill
  \begin{minipage}{0.48\textwidth}
    \begin{align}
      \label{eq:wsys}
      \delta_{\rm sys} \bar{\mathcal{O}}^2 &= \sum_f w^f (\mathcal{O}^f - \bar{\mathcal{O}})^2, \\ 
      \label{eq:werr}
      \delta \bar{\mathcal{O}} &= \sqrt{\delta_{\rm stat} \bar{\mathcal{O}}^2 + \delta_{\rm sys} \bar{\mathcal{O}}^2}.
    \end{align}  
  \end{minipage}
\end{subequations} 

\begin{figure}[t]
    \centering
    \includegraphics[width=.75\textwidth]{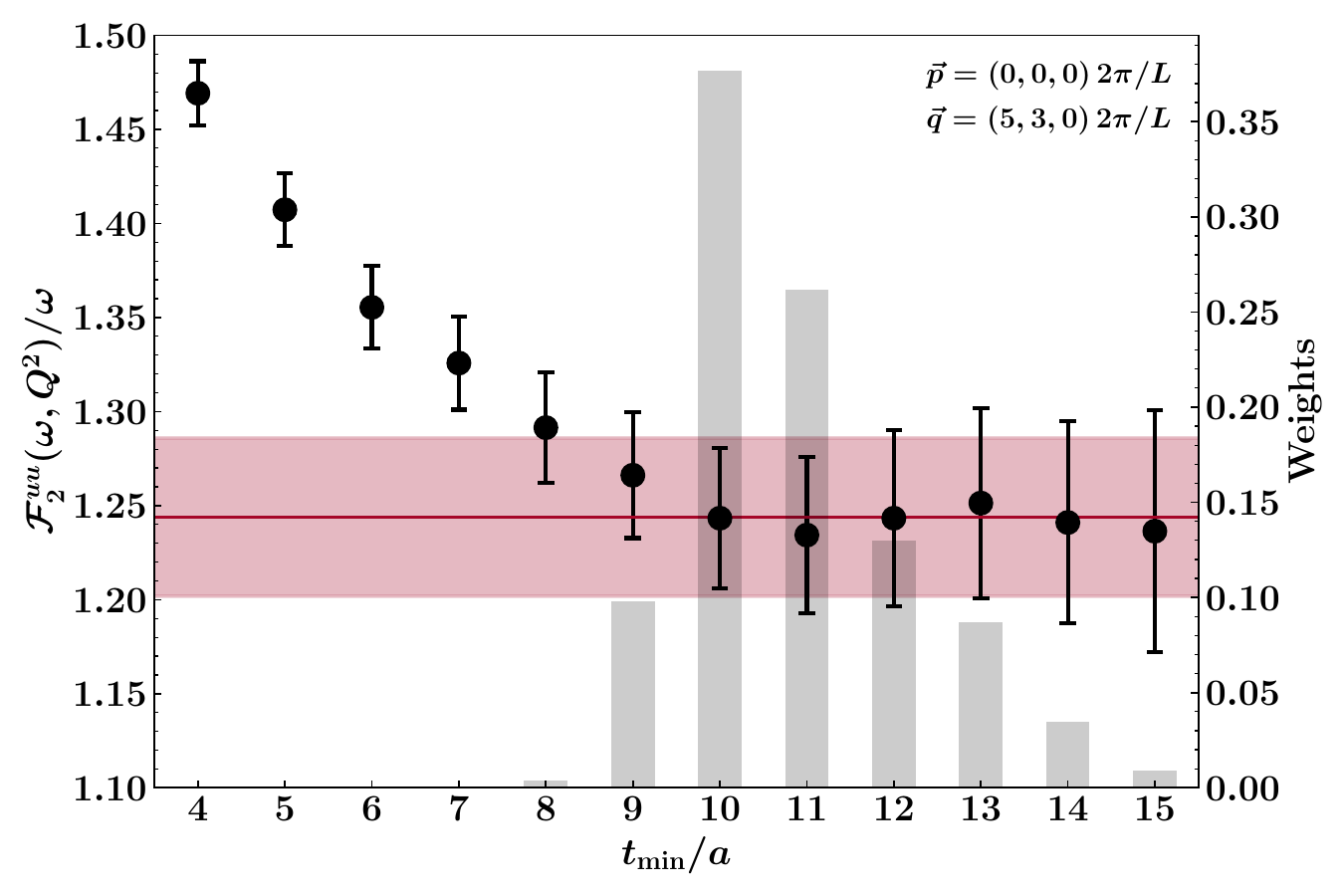}
    \caption{\label{fig:ooo_F2_uu_q530}The $\mathcal{F}_2$ Compton structure function obtained via \Cref{eq:compF2} for $(\pb, \qb) = ((0,0,0) \latmom, (5,3,0) \latmom)$, where $L=48a$. The $uu$ contribution determined on the $SU(3)$ symmetric ensemble is shown. $t_{\rm max}=20a$ for this case and we dropped the $t_{\rm min} > 15a$ points for clarity due to their vanishing weights. The grey bars associated with the right y-axis denote the weight, \Cref{eq:weight}, of each point. The final estimate, \Cref{eq:wavg}, is shown by the solid line, and the inner and outer (barely visible) shaded bands denote the statistical, \Cref{eq:wstat}, which is dominant, and the total, \Cref{eq:werr}, uncertainties.
    }
\end{figure}

We typically compute the energy shifts $\Delta E_{N_\lambda}(\pb, \qb)$, for two $|\lambda|$ values and perform fits polynomial in $\lambda$ (\Cref{eq:enshift_oo}), where the higher-order terms are neglected, to determine the Compton amplitude for each fit window, $f$, within their respective bootstrap samples. It has been shown that the higher order $\mathcal{O}(\lambda^4)$ terms are heavily suppressed for $|\lambda| = \mathcal{O}(10^{-2})$~\cite{QCDSFUKQCDCSSM:2022ncb,Can:2025zsr}. We show a representative $\mathcal{F}_2$ Compton structure function for a nucleon at rest, i.e. $\omega=0$, in \Cref{fig:ooo_F2_uu_q530} as calculated using the energy shifts extracted on several fit windows, along with the weight of each estimate. 

We follow the procedure outlined above to extract the $\mathcal{F}_2$ Compton structure function for a range of $\omega$ values for each $\qb$ we consider for the $uu$ and $dd$ contributions. In order to propagate the uncertainty estimates obtained from the weighted averaging method, we pick a fit window that gives a conservative estimate of the total uncertainty and an accurate central value for each $(\pb, \qb)$ pair. For instance, we pick $t_{\rm min} = 12a$ for the case given in \Cref{fig:ooo_F2_uu_q530}, as opposed to $t_{\rm min} = 10a$ which, although has a larger weight, underestimates the total uncertainty.

Extraction of the moments from the Compton structure functions follows the methodology described in~\cite{PhysRevD.102.114505}. A fit to $\mathcal{F}_2/\omega$ (\Cref{eq:ope_moments2}) is performed in a Bayesian framework to determine the lowest even Mellin moment of the $F_2$ structure function. The series (\Cref{eq:ope_moments2}) is truncated at $n=6$ (inclusive) when determining the moments. The consecutive moments are enforced to be positive definite and monotonically decreasing. The sequences of individual $uu$ or $dd$ moments are selected according to a multivariate probability distribution, $\operatorname{exp}(-\chi^2/2)$, where,
\begin{equation}\label{eq:chi2}
  \chi^2 = \sum_{i,j} \left[ \overline{\mathcal{F}}_{2,i} - \overline{\mathcal{F}}_2^\text{obs}(\omega_i) \right] C^{-1}_{ij} \left[ \overline{\mathcal{F}}_{2,j} - \overline{\mathcal{F}}_2^\text{obs}(\omega_j) \right],
\end{equation} 
is the $\chi^2$ function with the covariance matrix $C_{ij}$, ensuring the correlations between the points are taken into account.
\begin{figure}[t]
    \centering
    \includegraphics[width=.725\textwidth]{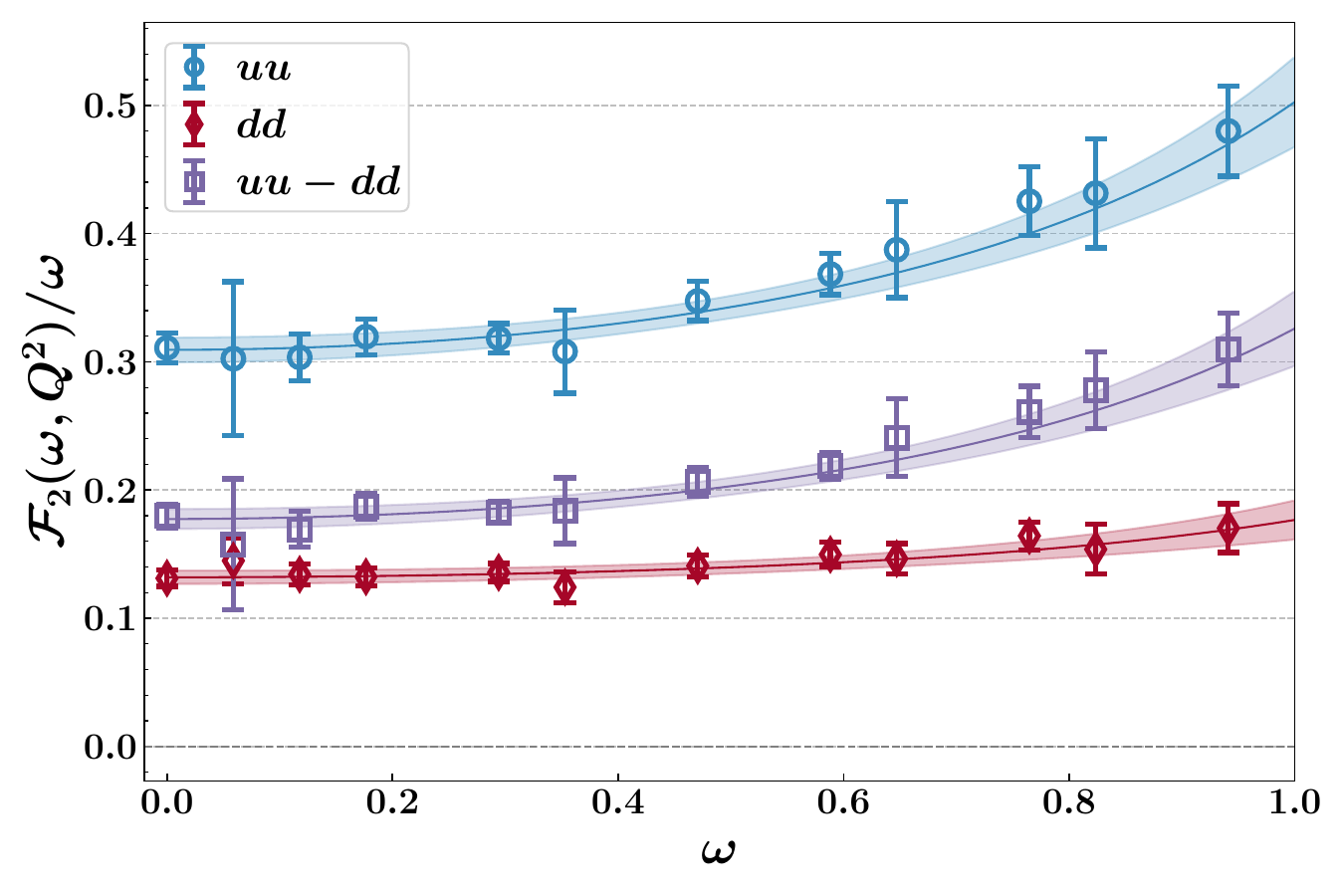}
    \caption{\label{fig:F2_q530_moments}$\omega$ dependence of the Compton structure function $\mathcal{F}_2$ at $Q^2 \sim 5 \, {\rm GeV}^2$. We show the $uu$ (blue circles) and $dd$ (red diamonds) contributions and the constructed $uu-dd$ (purple squares) results. Coloured shaded bands show the fits with their 68\% credible region of the highest posterior density. We stress that we do not fit to the $uu-dd$ points but show the constructed points ($\mathcal{F}_2^{uu}(\omega)/\omega - \mathcal{F}_2^{dd}(\omega)/\omega$) and fit band $\left( \sum_{n=1}^6 \omega^{2n-2} [M^{(2)}_{2n, uu}-M^{(2)}_{2n, dd}](Q^2) \right)$.
    }
\end{figure}  
Note that the monotonic decreasing condition is not necessarily true for the isovector, $uu-dd$, moments. Therefore, the Bayesian priors for the $uu$ and $dd$ are treated independently. However, by sampling the $uu$ and the $dd$ points within the same trajectory, we ensure underlying correlations between those points are accounted for. Hence, the indices $i$, $j$ in \Cref{eq:chi2} run through all the $\omega$ values and both flavours. The isovector $uu-dd$ moments are constructed via, $M^{(2)}_{2n, uu-dd}(Q^2) = M^{(2)}_{2n, uu}(Q^2)-M^{(2)}_{2n, dd}(Q^2)$. Fits depicting the extraction of the $uu$ and $dd$ moments are shown in \Cref{fig:F2_q530_moments}, along with the constructed $uu-dd$ moments. We find that the intercepts of the fit curves and the $\omega=0$ results to be in very good agreement. The monotonic increase of the points are well-described by our truncated series.

\section{Results and discussion} \label{sec:res}
\begin{figure}[t]
  \centering
  \includegraphics[width=.65\textwidth]{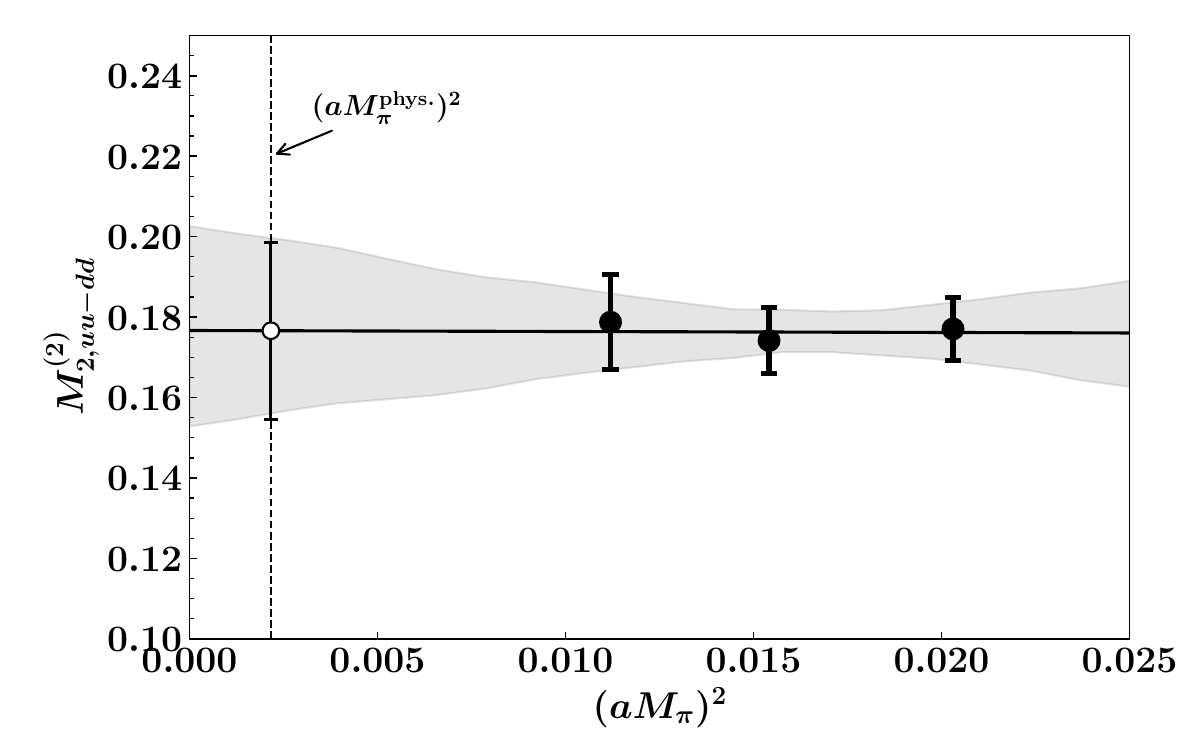}
  \caption{\label{fig:m2_umd_chiral}The lowest even isovector moment, $M_{2, uu-dd}^{(2)}(Q^2)$ as a function of $am_\pi^2$ along the $\overline{m} = {\rm constant}$ line at fixed $Q^2 \sim 5 \, {\rm GeV}^2$. Solid line and the shaded band show the fit curve and the $1\sigma$ uncertainty respectively. See the main text for the fit function.
  }
\end{figure}

We focus on our results for the lowest even isovector moment, $M^{(2)}_{2, uu-dd}(Q^2)$, in this section. In order to study the quark-mass dependence of the isovector moment, we carry out the analysis described in \Cref{sec:simu} at fixed $Q^2 \sim 5 \; {\rm GeV}^2$ on three ensembles that have varying $m_\pi$ starting from the $SU(3)$ point and decreasing down to $m_\pi \approx 300 \; {\rm MeV}$ along the $\overline{m} = {\rm constant}$ line. We show the extracted moments as a function of $(a m_\pi)^2$ in \Cref{fig:m2_umd_chiral}, along with a fit (solid band) linear in $(a m_\pi)^2$, $f(a m_\pi^2) = M_{2, uu-dd}^{(2), {\rm phys.}} + c \, (a m_\pi)^2$. Our lattice moments show almost no dependence to the quark mass in the range we have considered. We find 
\begin{equation} \label{eq:m2_mom}
  M_{2, uu-dd}^{(2), {\rm phys.}}(Q^2 \sim 5 \; {\rm GeV}^2) = 0.177(22),
\end{equation}
where the $\sim 10\%$ uncertainty quantifies the statistical and one source of the systematic uncertainties. At $Q^2 \sim 5 \; {\rm GeV}^2$, the leading contribution to the moment \Cref{eq:m2_mom} is dominated by the twist-2 operator. The quark mass dependence of the matrix element of the twist-2 operator on the $\overline{m} = \textrm{constant}$ line has been worked out in Ref.~\cite{Bickerton:2019nyz}. In our case the expansion coefficients are constrained by the fact that $\langle x \rangle_u + \langle x \rangle_d + \langle x \rangle_s + \langle x \rangle_g = 1$ is independent of the quark masses. If singlet and nonsinglet contributions are unrelated, we thus would expect no mass dependence. Controlling the finite volume and lattice discretisation errors would require further calculations on varying volumes and lattice spacings, which we leave to a future work, however, note that calculations by other groups~\cite{Mondal:2020ela, Djukanovic:2024krw} show only a mild dependence to lattice spacing and finite volume for the leading-twist part of this quantity.
\begin{figure}[t]
  \centering
  \includegraphics[width=.75\textwidth]{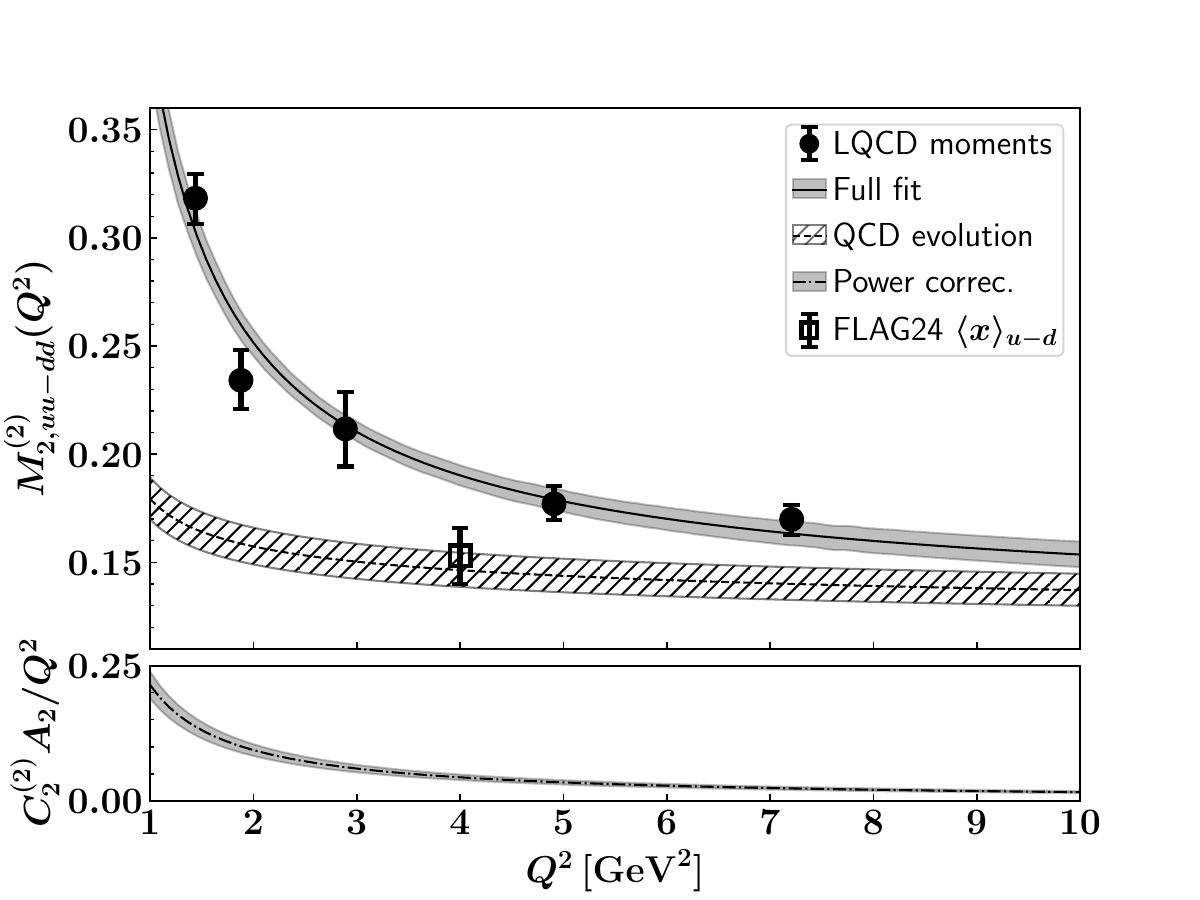}
  \caption{\label{fig:m2_umd_running}The lowest even isovector Mellin moments (LQCD moments) of the $F_2$ structure function as a function of $Q^2$ extracted from our lattice $\mathcal{F}_2$ Compton structure functions. The solid band (Full fit) shows our fit to the LQCD moments using \Cref{eq:nnlo_m2}, the hatched band (QCD evolution) is the NNLO evolution of the leading-twist moment, and the bottom panel shows the contribution of the higher-twist term. The FLAG average for the leading-twist moment~\cite{FLAG:2024oxs} is indicated by the open square.   
  }
\end{figure}

The operator product expansion relates the structure function moments to the PDF moments in the $Q^2 \to \infty$ limit as given in \Cref{eq:leading-twist}. Utilising the structure function moments we have calculated in the range $1 \lesssim Q^2 \lesssim 7.5 \; {\rm GeV}^2$ at the $SU(3)$ symmetric point $(m_\pi \approx 410 \; {\rm MeV})$, we extract the leading-twist contribution, while estimating the leading higher-twist part. The running of the lowest non-singlet leading-twist moment is known to next-to-next-leading order (NNLO)~\cite{Larin:1993vu}, and we adopt a multiplicative prescription assuming the higher-twist contribution shares the leading-twist Wilson coefficient. Therefore we have,
\begin{equation} \label{eq:nnlo_m2}
  M^{(2)}_{2,uu-dd}(Q^2) = C_{2,uu-dd}^{(2)}\left(\frac{Q^2}{\mu^2},g(\mu)\right) \left[ v_{2}^{uu-dd}(\mu) + \frac{A_2^{uu-dd}(\mu)}{Q^2} \right],
\end{equation}
where $\mu$ is the renormalisation scale, $v_{2}^{uu-dd}(\mu)$ is the leading-twist matrix element corresponding to the PDF moment $\langle x \rangle_{u-d}$, and $A_2^{uu-dd}(\mu)$ is the twist-4 matrix element. Here, 
\begin{equation}
    C_{2,uu-dd}^{(2)}\left(\frac{Q^2}{\mu^2},g(\mu)\right) = \left( \frac{a_s(Q^2)}{a_s(\mu^2)} \right)^{\gamma^{(0)}/2\beta_0}\left[ 1 
    + 158.07 \, \frac{a_s(Q^2)}{\beta_0^2} 
    + 58411.28 \, \frac{a_s^2(Q^2)}{\beta_0^4}  \right],
\end{equation}
is the Wilson coefficient calculated to NNLO~\cite{Larin:1993vu} in the ${\rm \overline{MS}}$ scheme, with $a_s = \frac{\alpha_s}{4\pi} = \frac{g^2}{16\pi}$, $\gamma^{(0)}$ the leading-order coefficient of the anomalous dimensions, $\beta_0 = 11 - (2/3) n_f$ the 1-loop $\beta$-function coefficient, and $n_f$ the number of active flavours. We set the reference scale to $\mu^2 = 4 \; {\rm GeV}^2$ and take $n_f=3$ since we do not have a charm quark in our lattice calculations. The strong coupling constant $\alpha_s(Q^2)$ is estimated to NNLO accuracy in the ${\rm \overline{MS}}$ scheme following the PDG definition~\cite{ParticleDataGroup:2024cfk} by using the 3-flavour $\Lambda_{\rm \overline{MS}}^{(3)} = 338(10) \; {\rm MeV}$ quoted by the Flavour Lattice Averaging Group (FLAG)~\cite{FLAG:2024oxs}. 

Our calculated structure function moments (black points) are shown in \Cref{fig:m2_umd_running} as a function of $Q^2$, along with the fit curves for \Cref{eq:nnlo_m2} (solid band) and the NNLO evolution of the leading-twist moment, \Cref{eq:nnlo_m2} without the higher-twist term (hatched band). We find,
\begin{equation}
  \langle x \rangle_{u-d} = 0.141(8),
\end{equation}
at the reference scale $\mu^2 = 4 \; {\rm GeV}^2$, in good agreement with the $N_f = 2+1$ average quoted by FLAG~\cite{FLAG:2024oxs}. 

The bottom plot in \Cref{fig:m2_umd_running} shows the higher-twist contribution as a function of $Q^2$, and indicates that the higher-twist term is significant even at the typical scales ($Q^2 \sim 4 \; {\rm GeV}^2$) where the leading-twist approximation is assumed to be valid in global QCD analyses. Our result shows that the higher-twist term contributes as much as $30\%$ of the leading-twist moment at $Q^2 = 4 \; {\rm GeV}^2$ to the physical structure function moment, and the leading-twist part does not dominate even at $Q^2 \sim 10 \; {\rm GeV}^2$, where the higher-twist term gives an approximate $10\%$ contribution. 

Although there remains the usual lattice systematic errors (e.g., finite-volume, discretisation, and unphysical quark masses) to quantify, we remind the reader that calculations of the leading-twist isovector moment by other groups~\cite{Mondal:2020ela, Djukanovic:2024krw} only show a mild dependence to lattice spacing and finite volume, and our results shown in \Cref{fig:m2_umd_chiral} indicate a very mild quark mass dependence for the isovector moment, hence we do not expect a drastic change of our quoted values that would invalidate the insights we draw from our results.

\section{Summary and outlook} \label{sec:sum}
We have calculated the lowest even isovector moment of the $F_2$ structure function in $2+1$ flavour lattice QCD with varying quark masses corresponding to $m_\pi \approx [410, 360, 300] \; {\rm MeV}$, at fixed volume $V = 48^3 \times 96$ and coupling $\beta = 5.65$ corresponding to a lattice spacing $a = 0.068(3) \, {\rm fm}$. Our calculation utilised the established Feynman-Hellmann approach to directly compute the physical Compton amplitude and extract the moments of the physical structure functions. We have presented our results for the quark-mass dependence of the lowest isovector moment, and estimated its value at the physical quark-mass point with $\sim 10\%$ uncertainty at fixed $Q^2$. By studying the $Q^2$ dependence of the moments, we have separated the leading- and higher-twist contribution to the physical moment and estimated the leading-twist matrix element, which is the PDF moment $\langle x \rangle_{u-d}$. Our $\langle x \rangle_{u-d}$ estimate shows a good agreement with the $2+1$-flavour average quoted by the Flavour Lattice Averaging Group's review~\cite{FLAG:2024oxs}. In this contribution, we have reported on our progress in quantifying one of the lattice systematic errors, e.g. unphysical quark masses, that affects our determination of the Compton amplitude and structure function moments. Next steps include calculations towards lighter quark masses, and with smaller lattice spacings and larger volumes to estimate a full error budget for our approach.

\acknowledgments
The numerical configuration generation (using the BQCD lattice QCD program~\cite{Haar:2017ubh})) and data analysis (using the Chroma software library~\cite{Edwards:2004sx}) was carried out on the Extreme Scaling Service (DiRAC, EPCC, Edinburgh, UK), the Data Intensive Service (DiRAC, CSD3, Cambridge, UK), the Gauss Centre for Supercomputing (NIC, Jülich, Germany), the NHR Alliance (Germany), and resources provided by the NCI National Facility in Canberra, Australia (supported by the Australian Commonwealth Government), the Pawsey Supercomputing Centre (supported by the Australian Commonwealth Government and the Government of Western Australia), and the Phoenix HPC service (University of Adelaide). RH is supported by STFC through grant ST/X000494/1. PELR is supported in part by the STFC under contract ST/G00062X/1. GS is supported by DFG Grant SCHI 179/8-1. KUC, RDY and JMZ are supported by the Australian Research Council grants DP190100297, DP220103098, and DP240102839. For the purpose of open access, the authors have applied a Creative Commons Attribution (CC BY) licence to any Author Accepted Manuscript version arising from this submission.

\providecommand{\href}[2]{#2}\begingroup\raggedright\endgroup

\end{document}